\documentclass[manuscript, screen, authorversion]{acmart}

\AtBeginDocument{%
  }

\copyrightyear{2025} 
\acmYear{2025} 
\setcopyright{rightsretained}
\acmConference[AutomotiveUI Adjunct '25]{17th International Conference on Automotive User Interfaces and Interactive Vehicular Applications}{September 21--25, 2025}{Brisbane, QLD, Australia}
\acmBooktitle{17th International Conference on Automotive User Interfaces and Interactive Vehicular Applications (AutomotiveUI Adjunct '25), September 21--25, 2025, Brisbane, QLD, Australia}
\acmDOI{10.1145/3744335.3758485}
\acmISBN{979-8-4007-2014-7/2025/09}

\acmSubmissionID{32}

\usepackage[T1]{fontenc}
\usepackage{xcolor}
\usepackage{xspace}
\usepackage[capitalise]{cleveref}
\usepackage{subcaption,amsfonts,dcolumn} % for subtable
\usepackage{gensymb} % for \degree

\newcommand{\ie}{\emph{i.e.}\xspace}

\newcommand{\LevelOne}{Level~1\xspace}
\newcommand{\LevelTwo}{Level~2\xspace}
\newcommand{\LevelThree}{Level~3\xspace}

\begin{document}

\title[Electrodermal Activity-Based Indicators of Driver Cognitive Distraction]{Are Electrodermal Activity-Based Indicators of Driver Cognitive Distraction Robust to Varying Traffic Conditions and Adaptive Cruise Control Use?}

\author{Ana{\"i}s Halin}
\email{anais.halin@uliege.be}
\orcid{0000-0003-3743-2969}
\affiliation{%
  \institution{University of Liège}
  \department{Department of Electrical Engineering and Computer Science}
  \city{Liège}
  \country{Belgium}
}

\author{Marc Van Droogenbroeck}
\email{m.vandroogenbroeck@uliege.be}
\orcid{0000-0001-6260-6487}
\affiliation{%
  \institution{University of Liège}
  \department{Department of Electrical Engineering and Computer Science}
  \city{Liège}
  \country{Belgium}
}

\author{Christel Devue}
\email{cdevue@uliege.be}
\orcid{0000-0001-7349-226X}
\affiliation{%
  \institution{University of Liège}
  \department{Department of Psychology}
  \city{Liège}
  \country{Belgium}
}

%\renewcommand{\shortauthors}{Halin et al.}

\begin{abstract}
In this simulator study, we investigate whether and how electrodermal activity (EDA) reflects driver cognitive distraction under varying traffic conditions and adaptive cruise control (ACC) use.
Participants drove in six scenarios, combining two levels of cognitive distraction (presence/absence of a mental calculation task) and three levels of driving environment complexity (different traffic conditions). Throughout the experiment, they were free to activate or deactivate ACC (ACC use, two levels).
We analyzed three EDA-based indicators of cognitive distraction: SCL (mean skin conductance level), SCR amplitude (mean amplitude of skin conductance responses), and SCR rate (rate of skin conductance responses).
Results indicate that all three indicators were significantly influenced by cognitive distraction and ACC use, while environment complexity influenced SCL and SCR amplitude, but not SCR rate. These findings suggest that EDA-based indicators reflect variations in drivers’ mental workload due not only to cognitive distraction, but also to driving environment and automation use.
\end{abstract}

\begin{CCSXML}
<ccs2012>
<concept>
<concept_id>10003120.10003121.10011748</concept_id>
<concept_desc>Human-centered computing~Empirical studies in HCI</concept_desc>
<concept_significance>500</concept_significance>
</concept>
<concept>
<concept_id>10003120.10003121.10003122.10011749</concept_id>
<concept_desc>Human-centered computing~Laboratory experiments</concept_desc>
<concept_significance>500</concept_significance>
</concept>
</ccs2012>
\end{CCSXML}

\ccsdesc[500]{Human-centered computing~Empirical studies in HCI}
\ccsdesc[500]{Human-centered computing~Laboratory experiments}

\keywords{Driver State, Cognitive Distraction, Workload, Indicator, Electrodermal Activity, EDA, Skin Conductance, Simulator Study, Driving Automation, Adaptive Cruise Control, Driving Environment Complexity, Traffic Conditions}

\maketitle

%%%%%%%%%%%%%%%%%%%%%%%% Introduction %%%%%%%%%%%%%%%%%%%%%%%%
\section{Introduction}
Until fully autonomous vehicles populate our roads entirely, ensuring that drivers are in an appropriate state to operate the vehicle or to supervise driving automation features remains critical for safety.
While many indicators have been extensively studied to characterize the driver state~\cite{Halin2021Survey}, further investigation is needed into how these indicators behave under varying traffic conditions and during the use of automation.
In this study, we focus on physiological indicators of driver cognitive distraction, specifically electrodermal activity (EDA). We analyzed three EDA-based indicators: (1)~the mean skin conductance level (SCL), (2)~the amplitude of skin conductance responses (SCR amplitude), and (3)~the rate of skin conductance responses (SCR rate), across six scenarios in a driving simulator.
These scenarios varied along two dimensions: (1)~driver state, with two levels of cognitive distraction (\ie, presence or absence of a secondary mental calculation task), and (2)~driving environment complexity, with three levels (\ie, increasing traffic density and the presence or absence of road construction restricting the number of traffic lanes). Additionally, participants were free to activate or deactivate the adaptive cruise control (ACC) at any time during any of the six scenarios.

The data analyzed in the present study were collected during a simulator study previously described in~\cite{Halin2025Effects}, which aimed to examine (1)~how driver cognitive distraction and driving environment influence ACC use, and (2)~how ACC use influences driving performance.
The current study investigates whether and how three indicators of EDA reflect driver cognitive distraction under varying traffic conditions and ACC use.

%%%%%%%%%%%%%%%%%%%%%%%% Related work %%%%%%%%%%%%%%%%%%%%%%%%
\section{Related Work}
Cognitive distraction is defined by the National Highway Traffic Safety Administration (NHTSA)~\cite{NHTSA2010Overview} as the mental workload associated with a task that involves thinking about something unrelated to the driving task. 
Monitoring cognitive distraction is, fundamentally, a monitoring of mental workload, which is affected not only by the complexity and demands of the driving task itself, but also by any concurrent secondary task (\ie, cognitive distraction).
For this reason, the present study aims to investigate not only how selected indicators reflect the driver’s cognitive state, but also how these indicators behave under varying traffic conditions and during the use of driving automation features.

A common approach to monitoring driver cognitive distraction is through physiological indicators, such as EDA~\cite{Yusoff2017Selection,Kajiwara2014Evaluation}, pupil diameter~\cite{Yokoyama2018Prediction}, and heart activity~\cite{Paxion2014Mental, Reimer2009AnOnroad}.
In the present study, we focus exclusively on EDA, as it can be collected with high reliability and is closely linked to sympathetic nervous system activity, which is known to respond to cognitive stimulation~\cite{Li2022Sensitivity}.

EDA refers to autonomic changes in the electrical properties of the skin, driven by sweat gland activity regulated by the sympathetic nervous system~\cite{braithwaite2015guide}.  
EDA is typically decomposed into (1)~a tonic component, reflecting slow, baseline shifts in arousal over time, and (2)~a phasic component, representing rapid, transient changes in response to discrete stimuli or events. The most commonly studied EDA signal is skin conductance (SC), which can be measured via skin-surface electrodes~\cite{VanDooren2012Emotional}. 
The measure of the tonic component is referred to as the skin conductance level (SCL), while the abrupt changes in the phasic component, called peaks, are referred to as skin conductance responses (SCRs). 

\citet{Li2022Sensitivity} analyzed seventeen features extracted from SC, SCL, and SCR under varying cognitive load and identified SCR rate as the most influential indicator of driver arousal, while also recommending SCR amplitude. \citet{Radhakrishnan2020Measuring} demonstrated that both automation and the environment significantly influence SCR rate. Similarly, \citet{Foy2018Mental} found that road type significantly affects mean SC and SCR rate. 
However, none of these studies examined the combined influence of cognitive distraction, driving environment complexity, and ACC use on the EDA-based indicators to assess their reliability in dynamic, real-world driving conditions. 

%%%%%%%%%%%%%%%%%%%%%%%% Research questions %%%%%%%%%%%%%%%%%%%%%%%%
\section{Research Questions}

In this study, we examined the robustness of EDA-based indicators of driver cognitive distraction across varying traffic conditions and driving automation use. Specifically, we aimed at answering the following overarching research question (RQ): \emph{Are electrodermal activity (EDA)-based indicators of driver cognitive distraction robust to changes in traffic conditions and ACC use?}

To answer this question, we investigated the three following sub-questions:
\begin{itemize}
    \item[\textbf{RQ1}] How is the mean skin conductance level (SCL) affected by cognitive distraction, traffic conditions, and ACC use? 
    \item[\textbf{RQ2}] How is the mean amplitude of skin conductance responses (SCR amplitude) affected by cognitive distraction, traffic conditions, and ACC use? 
    \item[\textbf{RQ3}] How is the rate of skin conductance responses (SCR rate, \ie, number of peaks per minute) affected by cognitive distraction, traffic conditions, and ACC use? 
\end{itemize}

%%%%%%%%%%%%%%%%%%%%%%%% User study %%%%%%%%%%%%%%%%%%%%%%%%
\section{User Study}
In the present study, we investigated the effects of a cognitively distracting task on EDA-based indicators under varying traffic conditions and ACC use. We used data collected during a driving simulator experiment designed to investigate how drivers’ cognitive state and driving environment complexity influence their reliance on driving automation features. Full details regarding the materials, experimental tasks, and procedure can be found in the paper of~\citet{Halin2025Effects}.

\subsection{Participants}
In total, $31$ participants were recruited, but $2$ participants did not complete the experiment due to simulator sickness. Besides, EDA data from one participant were of generally poor quality and identified as outliers, leading to the exclusion of all data from that individual. Therefore, the data from $N=28$ participants ($21$ male, $7$ female; mean age $= 31.57$, SD $= 10.88$ years) were available, as data from the $2$ withdrawn participants were entirely discarded for the analysis.
All participants, recruited via posters, word-to-mouth or social media, had a valid driver’s license. 
The study was approved by the ethics committee of the Faculty of Psychology, Logopedics and Educational Sciences of the University of Li{\`e}ge, Belgium, under the reference 2223-081.
Participants all provided informed and signed consent before taking part in the experiment.

\subsection{Apparatus and Materials}
\subsubsection{Driving Simulator}
The experiment was conducted in a driving simulator developed by AISIN Europe, running a customized version of CARLA~\cite{Dosovitskiy2017CARLA}, which is an open-source simulation environment based on Unreal Engine. The setup included three large $50$-inch curved screens, an adjustable car seat, and a Fanatec system comprising a steering wheel, gear shifter, and pedals.  
The vehicle was equipped with an intelligent adaptive cruise control system, which not only adapted to the speed of the preceding vehicle, but also automatically slowed down in curves. A custom software suite, developed by AISIN Europe, was used to design the test scenarios, execute the simulations, and log experimental data. Verbal responses of participants to the distraction task were recorded via a microphone. 

\subsubsection{BIOPAC MP160 System}
\begin{figure}
     \centering
     \includegraphics[width=.5\linewidth]{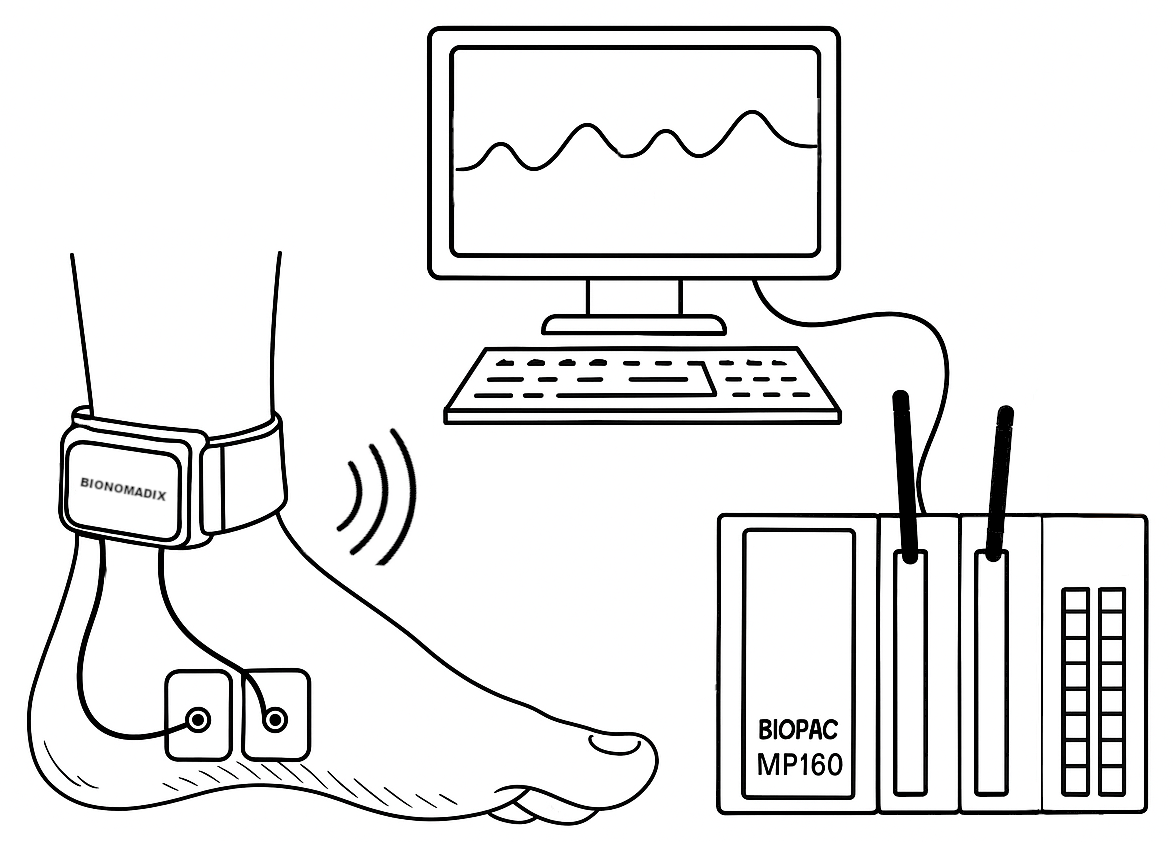}
     \Description{This image shows the placement of the two electrodes on the sole of the foot, as well as the setup with the BIOPAC to acquire the EDA signal.}
     \caption{Illustration of the setup for the acquisition of the EDA signal.}
     \label{fig:EDA_setup}
 \end{figure}

Participants were equipped with electrodes (BIOPAC EL507A) placed on their left foot to acquire and record EDA data using a BIOPAC MP160 system at a sampling rate of $2{,}000$ Hz (see \cref{fig:EDA_setup}). A gel (BIOPAC GEL101A) was applied to enhance conductivity between the skin and the electrode. 
To minimize driving inconvenience caused by the electrodes and cables, and to reduce motion artifacts, the electrodes were attached to the sole of the left foot, which rested on a designated area. Since the driving simulator replicated an automatic transmission vehicle, participants did not need to use their left foot to control the pedals. This electrodes' placement has been employed in previous studies~\cite{PunguMwange2022Measuring} and is known to provide reliable EDA measurements~\cite{VanDooren2012Emotional}.

\section{Procedure and Tasks}

We used a within-subjects driving simulator experiment in which participants completed six sessions of approximately eight minutes each. The route was identical across sessions and included urban and highway segments with varying traffic conditions. Participants were instructed to follow navigation cues, adhere to speed limits, and drive as they normally would, with the option to engage or disengage ACC at any time. 

Two experimental factors were manipulated: driving environment complexity (three levels, based on traffic density and road constructions) and cognitive distraction (presence or absence of a mental calculation task). 
\LevelOne of driving environments' complexity corresponded to low traffic density ($37$ other vehicles in the map). \LevelTwo involved medium traffic density ($80$ other vehicles), while \LevelThree maintained the same medium traffic density ($80$ other vehicles) but included three road construction zones that reduced road width to one or two lanes without causing traffic jams.
In the distraction condition, participants performed a secondary task consisting of responding aloud to arithmetic calculations delivered via a voice agent. Each calculation involved the addition of two two-digit numbers and occurred approximately three times per minute.

Participants were first equipped with EDA electrodes. Next, they filled out a pre-test questionnaire.
Participants were first familiarized with the set-up and the task. Then they performed six test scenarios in a counterbalanced order, with each level of complexity presented once with and once without the secondary task. At the beginning of each scenario, a brief synchronization procedure was conducted to align all recorded data streams. After each drive, participants completed a short questionnaire. For full details, see~\citet{Halin2025Effects}.
Finally, the experiment concluded with a five-minute relaxation period accompanied by soothing music.

\subsection{Measurements}  
Tonic (skin conductance level, SCL) and phasic (skin conductance response, SCR) components of the EDA signal were extracted using NeuroKit2 Python toolbox~\cite{Makowski2021Neurokit}. We computed the mean values of the SCL, the SCR amplitude (excluding the tonic component), and the SCR rate (number of peaks per minute). These indicators were extracted from the total duration of each session (\ie, approximately $8$ minutes).

%%%%%%%%%%%%%%%%%%%%%%%% Results %%%%%%%%%%%%%%%%%%%%%%%%
\section{Results}
Statistical analyses were conducted using JASP (Version 0.19.3)~\cite{JASP2025}. 
We report descriptive and inferential statistics separately for each indicator. Results were considered statistically significant at $p<.05$.

Each of the $28$ participants completed the $6$ scenarios, resulting from the combination of $3$ levels of driving environment complexity and $2$ levels of cognitive distraction. Within each scenario, they were free to activate and deactivate ACC as often as they wished. This design was expected to yield $28 \times 6 \times 2 = 336$ observations (including data from both ACC on and ACC off conditions). However, $4$ observations were missing because participants never engaged the ACC during an entire given scenario. These missing values resulted in unbalanced data when analyzing the effect of ACC use on the different indicators. Additionally, electrodermal data from one participant were identified as outliers for the \LevelTwo{} driving environment complexity with no distraction conditions using a Z-score threshold of $3$ and were thus discarded. 
As a result, the final analysis was conducted on $330$ observations from $28$ participants. A linear mixed model was fitted using Restricted Maximum Likelihood (REML) estimation and Satterthwaite’s approximation for degrees of freedom. The model included driving environment complexity, cognitive distraction, and ACC as fixed effects, with a random intercept for participants to account for individual variability.

\subsection{EDA-Related Indicators of Cognitive Distraction (\textbf{RQ})}
\subsubsection{SCL (\textbf{RQ1})}
The linear mixed model analysis revealed significant main effects of cognitive distraction ($F(1,291.01) = 11.982, p<.001$), driving environment complexity ($F(2,291.01)=7.212, p<.001$), and ACC use ($F(1,291.00)=3.893, p=.049$) on SCL. There was no significant interaction between cognitive distraction and driving environment complexity ($F(2,291.01)=.160, p=.852$), cognitive distraction and ACC use ($F(1,291.01)=.170, p=.680$), driving environment complexity and ACC use ($F(2,291.01)=.123, p=.884$), and finally cognitive distraction, driving environment complexity and ACC use ($F(2,291.01)=.118, p=.889$).

\Cref{tab:SCL} indicates that SCL was higher when participants were cognitively distracted compared to when there were not, and that SCL was lower when ACC was engaged compared to when it was disengaged. 
Post-hoc pairwise comparisons (using contrasts) were conducted to explore the effect of driving environment complexity on SCL, averaging over distraction and ACC use. P-values are adjusted using Bonferroni correction. 
Results showed that SCL was significantly lower at $\LevelThree$ compared to both $\LevelOne$ ($b=1.933, SE=.569, z=3.399, p=.002$) and $\LevelTwo$ ($b=1.821, SE=.575, z=3.166, p=.005$) of driving environment complexity. Thus, SCL decreases as the driving environment complexity increases. 
No significant difference was however observed between $\LevelOne$ and $\LevelTwo$ ($b=.112, SE=.572, z=.196, p=1.0$).

\begin{table*}[ht]
    \centering
    \caption{Averaged EDA-related indicators of cognitive distraction (SD; Min-Max) for the $N=28$ participants when the ACC was engaged and disengaged across the six scenarios, \ie, in the three levels of driving environment complexity (DEC), with and without cognitive distraction.}
    \label{tab:EDA_indicators}

    \begin{subtable}{\textwidth}
    \centering
    \caption{Skin conductance level (SCL)}
    \label{tab:SCL}
    \begin{tabular}{l|c|c|c|c|}
        \cline{2-5}
                                                & \multicolumn{2}{c|}{\textbf{No distraction}}                & \multicolumn{2}{c|}{\textbf{Distraction}}\\
        \hline
        \multicolumn{1}{|l|}{\textbf{DEC} }     & \textbf{ACC Off}            & \textbf{ACC On}               & \textbf{ACC Off}             & \textbf{ACC On} \\    
        \hline
        \multicolumn{1}{|l|}{\textbf{\LevelOne (low)}}      & $6.3$ ($3.3$; $2.1$-$13.3$)   & $6.1$ ($3.3$; $1.6$-$13.7$)    & $6.7$ ($3.5$; $1.9$-$13.7$)   & $6.4$ ($3.5$; $1.4$-$13.3$) \\
        \multicolumn{1}{|l|}{\textbf{\LevelTwo (medium)}}   & $6.4$ ($3.5$; $2.4$-$13.6$)   & $5.6$ ($3.1$; $1.6$-$12.5$)    & $6.7$ ($3.6$; $2.0$-$14.5$)   & $6.5$ ($3.6$; $1.5$-$14.0$) \\
        \multicolumn{1}{|l|}{\textbf{\LevelThree (high)}}   & $5.8$ ($3.1$; $1.8$-$13.1$)   & $5.7$ ($3.1$; $2.2$-$12.6$)    & $6.2$ ($3.5$; $2.1$-$14.5$)   & $6.2$ ($3.6$; $1.8$-$14.5$) \\
        \hline
    \end{tabular}
    \end{subtable} 

    \bigskip

    \begin{subtable}{\textwidth}
    \centering
    \caption{Amplitude of the skin conductance responses (SCR amplitude)}
    \label{tab:SCR_amplitude}
    \begin{tabular}{l|c|c|c|c|}
        \cline{2-5}
                                                & \multicolumn{2}{c|}{\textbf{No distraction}}                & \multicolumn{2}{c|}{\textbf{Distraction}}\\
        \hline
        \multicolumn{1}{|l|}{\textbf{DEC} }     & \textbf{ACC Off}            & \textbf{ACC On}               & \textbf{ACC Off}             & \textbf{ACC On} \\    
        \hline
        \multicolumn{1}{|l|}{\textbf{\LevelOne (low)}}      & $.16$ ($.14$; $.005$-$.59$)   & $.14$ ($.15$; $.01$-$.48$)    & $.22$ ($.21$; $.01$-$.93$)   & $.16$ ($.14$; $.01$-$.52$) \\
        \multicolumn{1}{|l|}{\textbf{\LevelTwo (medium)}}   & $.20$ ($.20$; $.01$-$.089$)   & $.11$ ($.13$; $.01$-$.57$)    & $.23$ ($.21$; $.01$-$.71$)   & $.16$ ($.15$; $.01$-$.64$) \\
        \multicolumn{1}{|l|}{\textbf{\LevelThree (high)}}   & $.22$ ($.22$; $.01$-$.84$)    & $.18$ ($.23$; $.01$-$.87$)    & $.22$ ($.25$; $.01$-$1.13$)  & $.21$ ($.21$; $.01$-$.72$) \\
        \hline
    \end{tabular}
    \end{subtable} 

    \bigskip

    \begin{subtable}{\textwidth}
    \centering
    \caption{Rate of the skin conductance responses (SCR rate)}
    \label{tab:SCR_rate}
    \begin{tabular}{l|c|c|c|c|}
        \cline{2-5}
                                                & \multicolumn{2}{c|}{\textbf{No distraction}}                & \multicolumn{2}{c|}{\textbf{Distraction}}\\
        \hline
        \multicolumn{1}{|l|}{\textbf{DEC} }     & \textbf{ACC Off}            & \textbf{ACC On}               & \textbf{ACC Off}             & \textbf{ACC On} \\    
        \hline
        \multicolumn{1}{|l|}{\textbf{\LevelOne (low)}}      & $5.7$ ($1.9$; $2.6$-$10.5$)   & $4.5$ ($1.8$; $0.8$-$7.9$)    & $5.8$ ($2.0$; $2.4$-$10.4$)   & $5.4$ ($2.7$; $1.8$-$13.3$) \\
        \multicolumn{1}{|l|}{\textbf{\LevelTwo (medium)}}   & $5.7$ ($2.1$; $1.7$-$10.4$)   & $4.5$ ($2.2$; $1.1$-$10.0$)   & $5.4$ ($1.7$; $2.2$-$9.4$)    & $4.8$ ($1.8$; $2.6$-$9.1$) \\
        \multicolumn{1}{|l|}{\textbf{\LevelThree (high)}}   & $5.0$ ($2.4$; $1.3$-$12.7$)   & $4.7$ ($2.6$; $.98$-$11.5$)   & $5.7$ ($2.4$; $2.8$-$11.7$)   & $5.5$ ($2.8$; $2.4$-$12.5$) \\
        \hline
    \end{tabular}
    \end{subtable}  
\end{table*}

\subsubsection{SCR Amplitude (\textbf{RQ2})}
There were significant main effects of cognitive distraction ($F(1,291.03)=7.585, p=.006$), driving environment complexity ($F(2,291.03)=3.198, p=.042$), and ACC use ($F(1,291.00)=17.304, p<.001$) on SCR amplitude, but no significant interaction between cognitive distraction and driving environment complexity ($F(2,291.03)=.626, p=.535$), cognitive distraction and ACC use ($F(1,291.01)=.087, p=.769$), driving environment complexity and ACC use ($F(2,291.01)=1.603, p=.203$), and finally cognitive distraction, driving environment complexity and ACC use ($F(2,291.01)=1.046, p=.353$).

\Cref{tab:SCR_amplitude} indicates that SCR amplitude was higher when participants were cognitively distracted compared to when there were not. SCR amplitude was lower when ACC was engaged compared to when it was disengaged.
Post-hoc pairwise comparisons showed that SCR amplitude was higher at $\LevelThree$ than at $\LevelOne$ of driving environment complexity ($b=-.126, SE=.052, z=-2.442, p=.044$). Thus, SCR amplitude increases as the driving environment complexity increases.  
No significant difference was however observed between $\LevelOne$ and $\LevelTwo$ ($b=-.033, SE=.052, z=-.627, p=1.0$), and $\LevelTwo$ and $\LevelThree$ ($b=-.094, SE=.052, z=-1.791, p=.220$).

\subsubsection{SCR Rate (\textbf{RQ3})}
There were significant main effects of cognitive distraction ($F(1,291.13)=6.303, p=.013$) and ACC use ($F(1,291.05)=15.323, p<.001$) on SCR rate. However, there was no significant effect of driving environment complexity ($F(2,291.13)=.492, p=.612$) on SCR rate and no significant interaction between cognitive distraction and driving environment complexity ($F(2,291.12)=1.713, p=.182$), cognitive distraction and ACC use ($F(1,291.06)=2.002, p=.158$), driving environment complexity and ACC use ($F(2,291.06)=.988, p=.373$), and finally cognitive distraction, driving environment complexity and ACC use ($F(2,291.06)=.317, p=.728$).

\Cref{tab:SCR_rate} indicates that the SCR rate was higher when participants were cognitively distracted compared to when they were not. 
Similarly to SCL and SCR amplitude, SCR rate was lower when ACC was engaged compared to when it was disengaged.

%%%%%%%%%%%%%%%%%%%%%%%% Discussion %%%%%%%%%%%%%%%%%%%%%%%%
\section{Discussion}
% Effects of cognitive distraction
We found that all three selected EDA-based indicators (SCL, SCR amplitude, and SCR rate) are significantly influenced by cognitive distraction. Specifically, all three indicators showed higher values when drivers were cognitively distracted compared to when they were not. 
While \citet{Li2022Sensitivity} reported no significant effect of cognitive distraction on SCL, our findings of increased SCR rate and SCR amplitude with a cognitive load are consistent with theirs.

% Effects of driving environment complexity
SCL and SCR amplitude were significantly influenced by driving environment complexity, while SCR rate was not. SCR amplitude increased when the level of complexity increases, while SCL was unexpectedly lower in the most complex environment ($\LevelThree$) compared to the least complex environment ($\LevelOne$). 
In a previous study, \citet{Radhakrishnan2020Measuring} compared two different driving environments (rural vs. urban) and found a higher SCR rate in the rural environment compared to the urban one and explained this by the higher speed limits, narrower roads and tighter curves associated with the rural environment.
Similarly, \citet{Foy2018Mental} reported that SCR rate was influenced by road types, such as city center multi-lane or suburban single-lane roads. Thus, although our study did not find a significant main effect of traffic conditions on SCR rate, prior research suggests that other aspects of driving environment complexity, beyond traffic density and construction zones, may influence this indicator. 

% Effects of automation / ACC
Finally, our study reveals that SCL, SCR amplitude, and SCR rate were all significantly influenced by the use of ACC. Their values were lower when ACC was engaged than when it was not. 
This is consistent with the findings of \citet{Radhakrishnan2020Measuring}, who compared automated drives to manual drives and found a higher SCR rate during manual driving. 

% Global discussion
Except for SCR rate, which was not significantly influenced by the complexity of the driving environment, all three EDA-based indicators were affected by cognitive distraction, ACC use, and driving environment. These results suggest that drivers’ mental workload varies with these factors, and that electrodermal activity signal can capture workload fluctuations related not only to cognitive distraction but also to traffic complexity and the use of driving automation features.
Moreover, the fact that all indicators showed lower values during ACC use supports the idea that increasing vehicle automation can help reduce drivers’ mental workload. 
Furthermore, these results highlight the potential of monitoring the state of drivers to implement adaptive automation systems (\ie, systems capable of adjusting their level of automation dynamically based on both the driver state and the driving environment). 
However, for such systems to become a reality, future work is needed to better understand the complex relationships between driver state, driving environment complexity, and driving automation, to determine when and how to adapt the level of automation, and to ensure that adaptation strategies enhance both safety and driver trust.

%%%%%%%%%%%%%%%%%%%%%%%% Conclusion %%%%%%%%%%%%%%%%%%%%%%%%
\section{Conclusion}
% Short summary of experiment and RQs
In this simulator study, we investigated how drivers' cognitive state (specifically, driver cognitive distraction), driving environment complexity (specifically, traffic conditions), and the use of driving automation features (specifically, ACC) influence EDA-based indicators (namely, SCL, SCR amplitude, and SCR rate).
The EDA data from $N = 28$ participants were analyzed. These participants drove in six different conditions, consisting of three levels of driving environment complexity and two levels of cognitive distraction, with the freedom to activate or deactivate ACC at any time during each scenario.  

% Results
Our results show that both cognitive distraction and ACC use have significant effects on all three EDA-based indicators, while driving environment complexity significantly impacts SCL and SCR amplitude, but not SCR rate. These findings suggest that EDA-based indicators are sensitive to fluctuations in drivers' mental workload, but that these fluctuations are not only resulting from cognitive distraction, but also from changes in the driving environment and the use of driving automation features. 

Moreover, the observed reduction in indicator values during ACC use supports the notion that increasing automation can lower drivers' mental workload. These results highlight the potential of driver state monitoring to support adaptive automation systems (\ie, systems capable of adjusting their level of automation dynamically based on both the driver state and the driving environment).

%%%%%%%%%%%%%%%%%%%%%%%% Acknowledgments %%%%%%%%%%%%%%%%%%%%%%%%
\begin{acks}
The work by A. Halin was supported by the SPW EER, Wallonia, Belgium under grant n°2010235 (ARIAC by \href{https://www.digitalwallonia.be/en/}{DIGITALWALLONIA4.AI}). 
The authors thank AISIN Europe for providing access to their driving simulator, and specifically acknowledge Richard Virlouvet and Frédéric Burguet for their assistance and support with the simulator.
\end{acks}

%%%%%%%%%%%%%%%%%%%%%%%% Bibliography %%%%%%%%%%%%%%%%%%%%%%%%
\bibliographystyle{ACM-Reference-Format}
\bibliography{main}

\end{document}